\documentclass{aa}
\input psfig.sty
\begin{document}

   \thesaurus{03     
              (03.20.8; 
               11.04.1; 
               11.05.2; 
               11.06.1)} 
   \title{The First VLT FORS1 spectra of Lyman-break candidates in the
HDF-S and AXAF Deep Field
\thanks{Based on observations
collected at the European Southern Observatory, Paranal, Chile}  
}
   \author{
	S.Cristiani\inst{1,2}
        \and
	I.Appenzeller\inst{3}
	\and
        S.Arnouts\inst{2,4}
	\and
	M.Nonino\inst{1,5}
	\and
	A. Arag\'on-Salamanca\inst{6}
	\and 
        C.Benoist\inst{4}
	\and
	L.da Costa\inst{4}
        \and
	M.Dennefeld\inst{5}
	\and 
	R.Rengelink\inst{4}
	\and 
	A.Renzini\inst{4}
	\and 
	T.Szeifert\inst{3}
        \and 
	S.White\inst{7}        
}

   \offprints{S.Cristiani (Stefano.Cristiani@eso.org)}

   \institute{
	{Space Telescope European Coordinating Facility,
        Karl-Schwarzschildstr. 2, 
        D-85748 Garching, 
        Germany}
   \and
	{Dipartimento di Astronomia dell'Universit\`a di Padova, 
	Vicolo dell'Osservatorio 5, 
	I-35122 Padova, Italy}
   \and
	Landessternwarte K\"onigstuhl, D-69117 Heidelberg, Germany
   \and 
        European Southern Observatory,  Karl-Schwarzschildstr. 2, 
        D-85740 Garching, Germany
   \and
	{Osservatorio Astronomico di Trieste, Via G.B. Tiepolo 11, 
40131 Trieste, Italy} 
   \and 
	Institut d'Astrophysique de Paris - CNRS, 98bis Boulevard 
Arago, F-75014 Paris, France 
   \and  School of Physics \& Astronomy, University of Nottingham, University Park,
 Nottingham NG7 2RD, U.K.
   \and Max-Planck Institut f\"ur Astrophysik, D-85740 Garching, Germany
             }

   \date{Received Feb 26, 2000; accepted \dots}
   \authorrunning {Cristiani et al.}
   \titlerunning {FORS1 spectra of Lyman-break candidates}
   \maketitle

   \begin{abstract}
    {We report on low-resolution multi-object spectroscopy 
    of 30 faint targets ($R \simeq 24-25$) in the
    HDF-S and AXAF deep field obtained with the VLT Focal Reducer/low
    dispersion Spectrograph (FORS1). Eight high-redshift
    galaxies with $2.75< z < 4$ have been identified. The spectroscopic
    redshifts are in good agreement with the photometric ones with a
    dispersion $\sigma_z = 0.07$ at $z<2$ and $\sigma_z = 0.16$ at
    $z>2$. The inferred star formation rates of the individual
    objects are moderate, ranging from a few to a few tens $M_{\odot}~
    {\rm yr}^{-1}$. Five out of the eight high-z objects do not show
    prominent emission lines. One object has a spectrum typical of an
    AGN. In the AXAF field two relatively 
    close pairs of galaxies have been identified,
    with separations of 8.7 and 3.1 proper Mpc and mean redshifts of
    3.11 and 3.93, respectively.
}

   \end{abstract}

%

\section{Introduction}
Observations of galaxies, now extending up to a redshift $z \sim 6$
(\cite{hu99}),
are starting to provide quantitative information on basic properties -
number densities, luminosities, colors, sizes, morphologies, star
formation rates (SFR), chemical abundances, dynamics and clustering -
over a large span of cosmic time.  These data are beginning to sketch
out a direct picture not only of the physical processes taking place
in the assembly of the first galaxies, but also of the formation and
evolution of large scale structure (LSS) from the primordial density
fluctuations.  Quantitative information is now available about the
evolution of the neutral hydrogen and metal content of the universe
since $z\simeq 4$, the galaxy
luminosity function since $z\simeq 1$,
the morphology of field and cluster galaxies since
$z\simeq 0.8$.
A recent dramatic addition to the general picture has been the
discovery of a large population of actively star-forming galaxies at
$z\simeq 3$ (\cite{SGDA96}, hereafter SGDA96).  
The ``Lyman break'' color-selection technique (\cite{SH92}) 
has proved a reliable and highly efficient method
to select galaxies in large numbers at $z\ga 2.5$, providing the first
opportunity for statistical studies of evolutionary processes in
galaxies beyond $z=1$.  Follow-up spectroscopy of the UV drop-out
candidates on the Keck telescopes shows most to lie in the expected
redshift range, $2.5 \leq z < 3.5$, with successful redshift
measurement for more than $70 \%$.  
The Lyman-break galaxies have spectra resembling those of nearby
starburst galaxies, are strongly clustered, with a co-moving correlation
length similar to present-day galaxies.  
SGDA96 inferred typical SFRs of $1-6
h^{-2} {\rm M}_{\odot} {\rm yr}^{-1}$ for their galaxies, 
assuming a critical density universe.
Dust corrections based on the UV continuum
slope and on near-IR spectroscopy of a few objects suggest values
larger by a mean factor of about 7 (\cite{pettini98}).  
From the width of saturated
interstellar absorption lines, SGDA96 inferred tentative 1D velocity
dispersions in the range $\sigma_{1D} = 180 - 320 {\rm km \, s}^{-1}$,
but Pettini et al. (1998) measure in the IR significantly narrower
line-widths $\sigma_{1D} = 55 - 190 {\rm km \, s}^{-1}$ for the Balmer
and [OIII] emission lines, albeit in a sample of only five objects.

A programme has been started with the ESO VLT to study systematically
galaxies at $z \sim 4$ with the aim to clarify the earliest phases of
the processes leading to the formation of galaxies and LSS, reaching a
redshift domain where observations are more cosmologically
discriminant (\cite{arnouts99}) and taking advantage of a wide
photometric coverage (in particular in the IR) to obtain mass
estimates of the detected objects.  We report here the results of
pilot observations carried out during the commissioning and the
science verification of the FORS1 instrument at the VLT-UT1.


\section{The photometric databases and the selection of the candidates}
Deep multicolor imaging of the HDF-S and AXAF deep field has been
obtained from HST and from the ground. In particular WFPC2 data,
consisting of deep images in the F300W, F450W, F606W and F814W
filters, cover an area of 4.7 sq.arcmin reaching $10 \sigma$ magnitude
limits of 26.8, 27.7, 28.2 and 27.7 (in a 0.2 sq.arcsec area,
\cite{williams99}).
$UBVRIJHK$ data over an area of 25 sq.arcmin, including the WFPC2
field, have been obtained at the ESO 3.5m New Technology Telescope
(NTT) as a part of the ESO Imaging Survey (EIS) program (\cite{EISHDF}).
They reach  $2\sigma$ limiting magnitudes of
$U_{AB}\sim 27$, $B_{AB}\sim 26.5$, $V_{AB}\sim 26$, $R_{AB}\sim 26$,
$I_{AB}\sim 25$, $J_{AB}\sim 25$, $H_{AB}\sim 24$ and $K_{AB}\sim
24$.
The EIS survey observed also the AXAF1 field (\cite{EISAXAF}).  25
sq.arcmin were covered in $UBVRIJK$ down to $U_{AB}\sim$ 27.0,
$B_{AB}\sim$ 27, $V_{AB}\sim$ 26.5, $R_{AB}\sim$ 26.5 and $I_{AB}\sim$
26, $J_{AB}\sim 24.5$ and $K_{AB}\sim$ 23.5.  Photometric catalogs
were derived from Lanzetta et al. (1999), da Costa et al. (1999) and
Benoist et al. (2000).  
Lyman-break galaxy candidates were selected by the EIS team on the
basis of two-color diagrams shortly after the EIS observations in
order to provide targets for the FORS1 commissioning and Science
Verification.
A more refined selection has been carried out after the observations
reported here on the basis of the photometric redshift technique
described by Arnouts et al. (1999).
\section{Spectroscopic Observations}

The present data have been retrieved from the ESO Public Archive.
Spectroscopic observations were carried out with the FORS1 instrument
(\cite{FORS1}) in multiple object spectroscopy (MOS) mode on December
1998 by the FORS1 Commissioning Team and on January 1999 for the FORS1 
Science Verification (see {\tt http://www.eso.org/science/ut1sv/}
and \cite{crisSV}).
In the FORS1 MOS mode 19 movable slit blade pairs can be placed in a
FOV of $6.8 \times 6.8$ sq.arcmin. The actual useful field in the
direction of the dispersion is somewhat less and depends on the length
of the spectra/dispersion. In the present case the Grism I150 was
used, providing a useful field of $3.5 \times 6.8$ sq.arcmin.
One configuration of slits was observed in the HDF-S and two in the AXAF1. 
The journal of the observations is given in Table~\ref{tab:obs}.
\begin{table}
\caption{Journal of the MOS Observations}
\begin{tabular}{lcccc}
\hline \hline
Field & $\alpha_{2000}$ & $\delta_{2000}$ & date &
exp.time (s) \\
\hline
HDF-S   & 22:32:46 &$-60$:34.1 & 1998-Dec-15 &1800 \\
HDF-S   & 22:32:46 &$-60$:34.1 & 1998-Dec-16 &3600 \\
HDF-S   & 22:32:46 &$-60$:34.1 & 1998-Dec-18 &1800 \\
HDF-S   & 22:32:46 &$-60$:34.1 & 1998-Dec-19 &1800 \\
HDF-S   & 22:32:46 &$-60$:34.1 & 1998-Dec-19 &1800 \\
HDF-S   & 22:32:46 &$-60$:34.1 & 1998-Dec-20 &1800 \\
HDF-S   & 22:32:46 &$-60$:34.1 & 1998-Dec-20 &1800 \\
AXAF1/A & 03:32:08 &$-27$:46.0 & 1999-Jan-16 &2100 \\
AXAF1/A & 03:32:08 &$-27$:46.0 & 1999-Jan-21 &3600 \\
AXAF1/A & 03:32:08 &$-27$:46.0 & 1999-Jan-21 &2200 \\
AXAF1/A & 03:32:08 &$-27$:46.0 & 1999-Jan-22 &2100 \\
AXAF1/B & 03:32:08 &$-27$:46.0 & 1999-Jan-18 &2100 \\
AXAF1/B & 03:32:08 &$-27$:46.0 & 1999-Jan-18 &2100 \\
AXAF1/B & 03:32:08 &$-27$:46.0 & 1999-Jan-19 &2100 \\
AXAF1/B & 03:32:08 &$-27$:46.0 & 1999-Jan-19 &1600 \\
AXAF1/B & 03:32:08 &$-27$:46.0 & 1999-Jan-22 &2100 \\
\hline
\hline
\label{tab:obs}
\end{tabular}
\end{table}
When no suitable candidate was available for the allowed range of
positions of a given slit, a random object in the field was chosen. 

The MOS observations were reduced within the MIDAS package, using
commands of the LONG and MOS contexts. For each object the available
2-D spectra were stacked and then an optimal extraction was carried
out.  
\begin{figure*}
\psfig{figure= 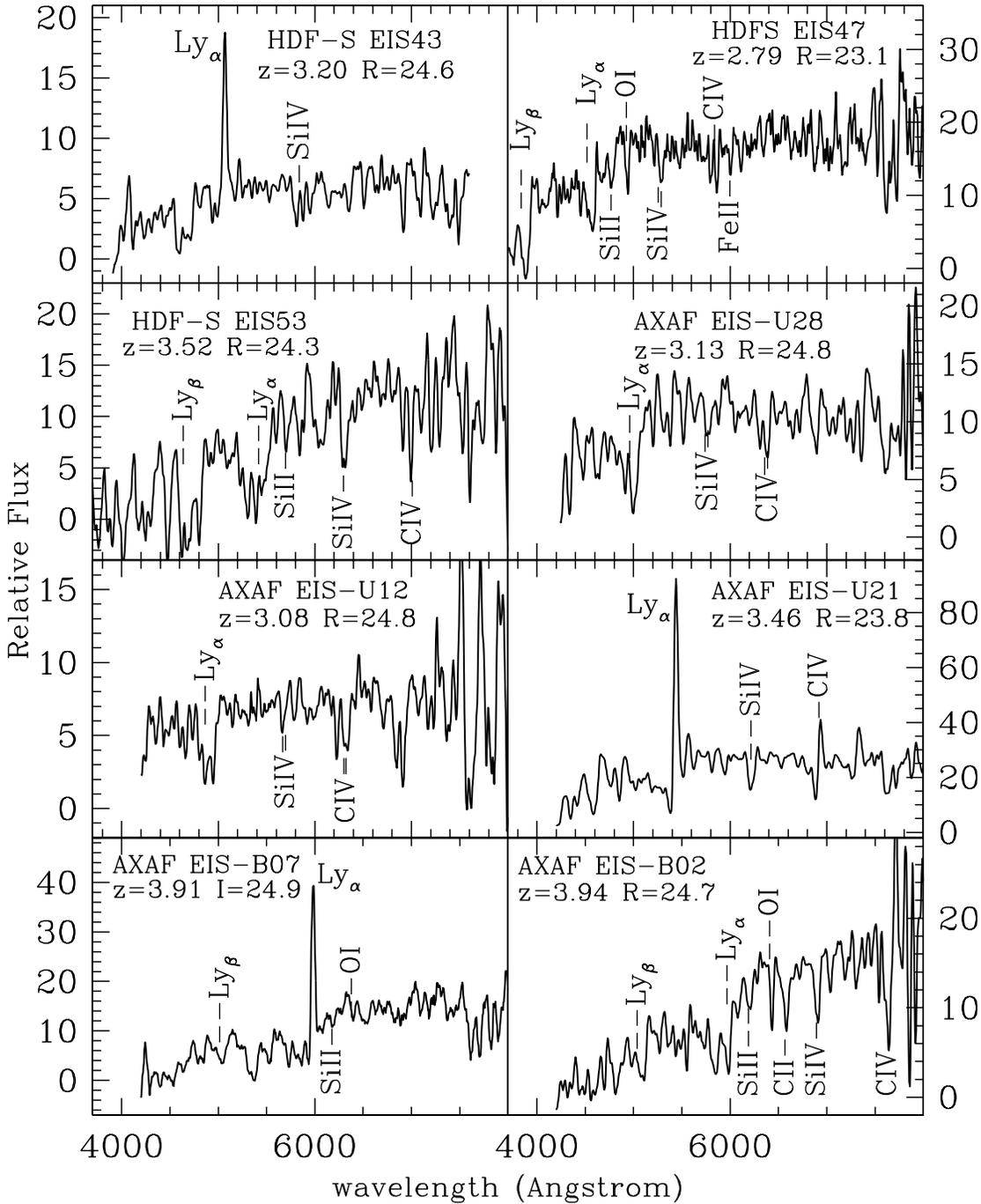,width=149mm}
\caption{Spectra of 8 high-redshift galaxies observed in the HDF-S and
AXAF deep field. The ordinate gives the relative flux density per Angstrom.}
\label{fig:spectra}
\end{figure*}
Tables 2 and 3 give the photometric information and the redshift (when
it has been possible to estimate it) for the objects observed in the
HDF-S and AXAF deep field, respectively.  
\begin{table}
\caption{Spectroscopic Identifications in the HDF-S}
\begin{tabular}{lccccc}
\hline \hline
Ident. & $\alpha_{2000}$ & $\delta_{2000}$ & z & $V_{AB}$ &$R_{AB}$\\
\hline
 EIS~18  & 22:32:30.9 & -60:32:44 &  -    & 25.18& 24.92\\
 EIS~23  & 22:32:31.9 & -60:35:16 &  -    & 25.25& 24.81\\
 ANON15  & 22:32:34.3 & -60:35:52 & Mstar & 23.52& 22.68\\
 ANON19  & 22:32:38.0 & -60:37:18 & 0.514 &   -  &   -  \\
 ANON18  & 22:32:38.7 & -60:37:02 & 0.410 &   -  &   -  \\
 EIS~33  & 22:32:40.0 & -60:36:21 &  -    & 24.36& 24.04\\
 EIS~36  & 22:32:42.2 & -60:34:46 & Mstar & $>26$& 25.67\\
 ANON02  & 22:32:45.0 & -60:30:55 & 0.514 &   -  &   -  \\
 ANON17  & 22:32:45.9 & -60:36:40 & 0.852 &   -  &   -  \\
 EIS~43  & 22:32:47.0 & -60:31:46 & 3.20\ & 24.52& 24.40\\
 ANON03  & 22:32:48.5 & -60:31:16 & 0.516 & 24.31& 23.58\\
 EIS~47  & 22:32:49.3 & -60:32:25 & 2.79\ &23.61$^\dag$&23.33$^\dag$\\
 ANON01  & 22:32:49.4 & -60:30:41 & 0.776 &   -  &   -  \\
 EIS~52  & 22:32:53.1 & -60:32:06 &  -    & 25.10& 24.49\\
 EIS~54  & 22:32:53.5 & -60:33:12 &  -    & 25.08& 24.86\\
 EIS~53  & 22:32:53.5 & -60:35:24 & 3.521 & 24.92& 24.24\\
 ANON09  & 22:32:53.7 & -60:33:37 & 0.565 & 22.52& 21.92\\
 EIS~58  & 22:32:54.7 & -60:34:31 &  -    & 24.61& 24.22\\
 EIS~60  & 22:32:55.4 & -60:33:55 & Mstar & 26.32& 25.30\\
\hline
\multicolumn{6}{l}
{$\dag$ complex morphology: photometry from Lanzetta
et al. 1999.}\\
\label{tab:HDFS}
\end{tabular}
\end{table}
\begin{table}
\caption{Spectroscopic Identifications in the AXAF Deep Field}
\begin{tabular}{lccccc}
\hline \hline
Ident. & $\alpha_{2000}$ & $\delta_{2000}$ & z & $V_{AB}$& $R_{AB}$\\
\hline
 ANON14  & 03:32:03.5 & -27:47:31 &  1.157 &  24.02 & 24.32 \\
 EIS~U28 & 03:32:03.6 & -27:43:40 &  3.132 &  24.79 & 24.87 \\
 EIS~U12 & 03:32:04.4 & -27:46:03 &  3.083 &  24.77 & 24.68 \\
 EIS~U21 & 03:32:05.0 & -27:44:32 &  3.462 &  23.65 & 23.74 \\
 EIS~B07 & 03:32:05.1 & -27:46:12 &  3.912 &  25.15 & 24.67 \\
 EIS~U01 & 03:32:05.8 & -27:48:16 &  star? &  23.64 & 22.98 \\
 EIS~B02 & 03:32:06.6 & -27:47:47 &  3.939 &  25.47 & 24.68 \\
 EIS~U14 & 03:32:09.1 & -27:45:35 &   -    &  25.37 & 24.95 \\
 EIS~B06 & 03:32:09.2 & -27:46:53 &  Mstar &  26.39 & 25.02 \\
 EIS~U19 & 03:32:09.6 & -27:45:14 &   -    &  25.48 & 25.02 \\
 EIS~B12 & 03:32:10.1 & -27:44:10 &   -    &  25.14 & 24.50 \\
\hline
\label{tab:AXAF}
\end{tabular}
\end{table}
In Column 1 the EIS
identifier refers to Lyman-break candidates in the original EIS lists
(for the AXAF field U- and B-dropouts are listed with the ``EIS U''
and ``EIS B'' prefix, respectively).  Random-chosen objects are listed
with the ``ANON'' prefix followed by the number of the slit in which
they were placed.
The spectra of 8 galaxies with redshifts between 2.8 and 4.0  
are shown in Fig.~\ref{fig:spectra}. 
These and more spectra are available in digital form at the URL
{\tt http://www.eso.org/science/ut1sv/MOS\_DR.html}.
In some cases (HDF-S: EIS~23, EIS~33, EIS~52; AXAF: EIS~U19, EIS~B12)
it was not possible to determine a redshift due to the lack of
significant spectral features
rather than to an insufficient S/N of the spectrum.
The photometric data of Columns 5 and 6 have been taken from the EIS
database in its most recent version (Arnouts, private communication). 
Five HDF-S ANON targets lie outside the EIS images and
no photometry is provided for them.
\section{Discussion}
At present, the spectroscopy of candidate Lyman-break galaxies has
been restricted to an area of 13.5 sq.arcmin in the HDF-S in which
$UBVRIJHK$ imaging is available and 25 sq.arcmin in the AXAF deep field
($22$ sq.arcmin covered in $UBVRIJK$ and 3 sq.arcmin in $UBVRI$ only).
{\it After} the FORS1 spectroscopic observations, which were based on
a preliminary list of Lyman-break candidates produced shortly after
the EIS imaging observations, we carried out a more
refined selection of galaxies with $z>2.75$ on the basis of a
photometric redshift code (described in Arnouts et al. 1999).
In the HDF-S $25$ candidates have been found down to a limiting mag
of $I_{AB} = 24.5$, while in the AXAF deep field $36$ 
candidates have been selected down to $R_{AB} = 25$.
Of the total $61$ candidates $10$ turned out to have
been observed during Commissioning and Science Verification: 
8 of them have been confirmed to be at high redshift, 2 resulted
in inconclusive spectra.
\begin{figure}
\psfig{figure=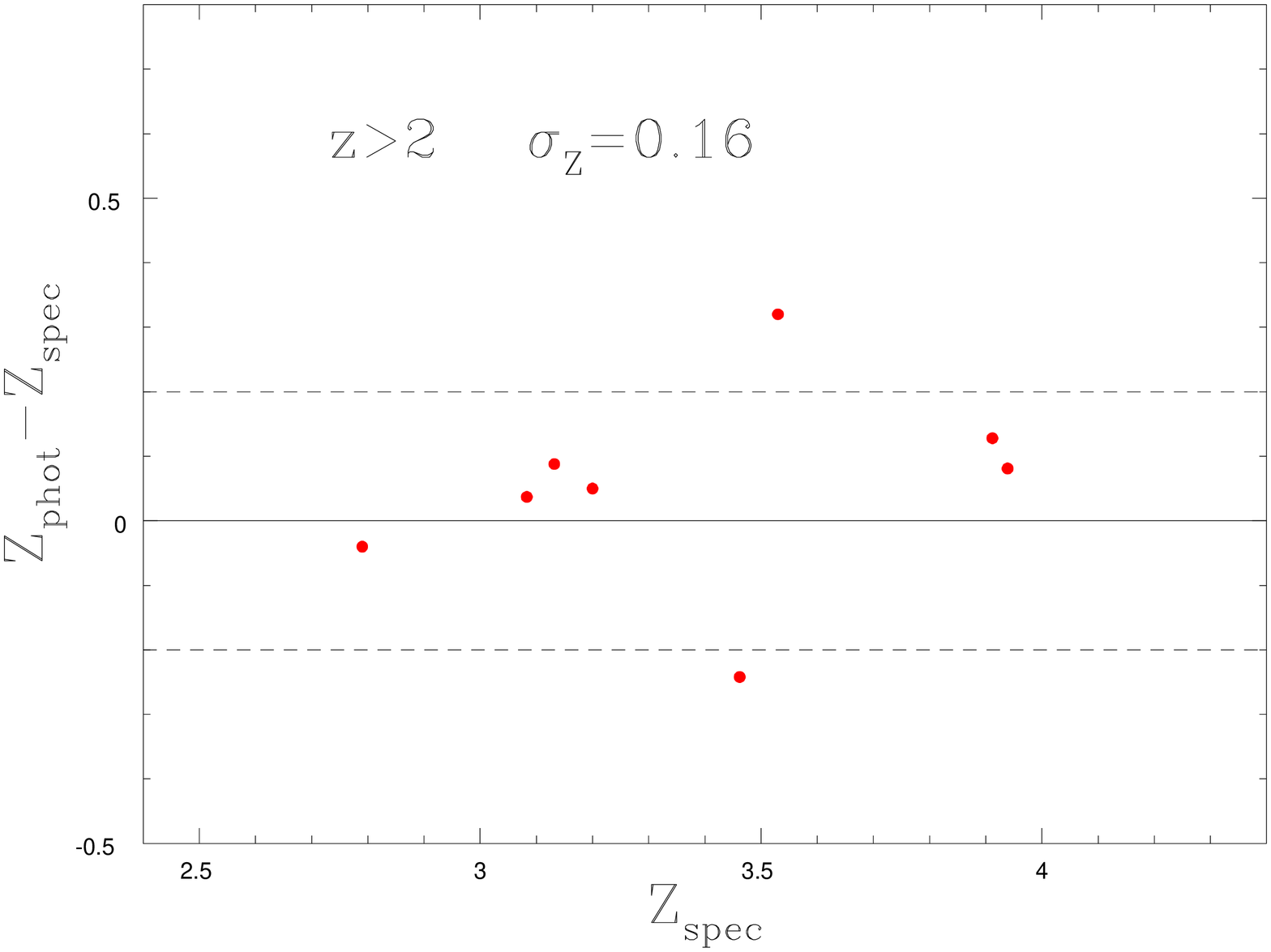,width=90mm}
\caption{Comparison of photometric and spectroscopic redshifts in the
HDF-S and AXAF deep field.}
\label{fig:DZ}
\end{figure}
Fig.~\ref{fig:DZ} shows the comparison between photometric and
spectroscopic redshifts for the 8 galaxies with $z>2.75$ observed so
far in the HDF-S and AXAF field. The resulting dispersion is $\sigma_z
(z>2.75) = 0.16$.  At lower redshift, including some preliminary
results in the HDF-S of Dennefeld et al. (2000, in preparation, see
{\tt http://www.iap.fr/hst/tmrresults.html}, the dispersion turns out
to be $\sigma_z (z<2)= 0.07$ (47 objects).

The properties of the high-z galaxies confirmed with the present
observations are summarized in Table~\ref{tab:properties}.
They have been inferred from the GISSEL models best-fitting the
photometric data (\cite{arnouts99}) imposing the redshift at the
spectroscopic value.
The Star formation rate (Column~2) is estimated with the UV continuum flux 
at 150 nm derived from the best fitting model.   
For a Salpeter IMF ($0.1 M_{\odot} < M < 125_{\odot}$) with constant SFR,
a galaxy with SFR $=1~ M_{\odot}$ yr $^{-1}$ produces 
$L(150$ nm $)=10^{40.15}$ erg ~ s$^{-1}$ \AA$^{-1}$
(\cite{madau96}).
Column~4 lists the SFR computed with the correction of the
intrinsic extinction (reported in Column~3), 
as obtained form the best-fit procedure. 
The Calzetti extinction law (\cite{calzetti97}) has been adopted.
Column~5 and 6 show the estimated age
\footnote{In this paper $H_o
=50$ and $q_o=0.5$ are assumed throughout.}
and stellar mass.
As typically found in surveys based on the ``Lyman-break'' technique,
the inferred star formation rates are moderate, ranging from a few to
a few tens $M_{\odot}$ yr$^{-1}$. Five out of eight high-z objects do not show
prominent emission lines. AXAF EIS-U21 has a spectrum typical of an
AGN, showing Ly-$\alpha$, CIV and possibly SiIV in emission 
with a P-Cyg profile.

\begin{table}
\caption{Properties of the Galaxies with $z>2.5$ in the HDF-S and AXAF fields.}
\begin{tabular}{lccccc}
\hline \hline
Identifier&SFR&$E_{BV}$&SFR&Age&Mass\\
&Uncorr.& &Corr.& & \\
&$M_{\odot}/$yr&mag&$M_{\odot}/$yr&Gyr&$\log(M_{\odot})$\\
\hline
HDF-EIS43& ~8 & 0.0 & ~8 & 0.7 & 10.6 \\
HDF-EIS47& 15 & 0.1 & 38 & 0.1 & 11.2 \\
HDF-EIS53& 12 & 0.1 & 19 & 0.1 & 10.8 \\
AX-EISU28& ~6 & 0.2 & 13 & 1.4 & 10.9 \\
AX-EISU12& ~6 & 0.0 & ~6 & 0.1 & 10.0 \\
AX-EISU21& AGN \\
AX-EISB07& 12 & 0.0 & 12 & 1.0 & 10.5 \\
AX-EISB02& 12 & 0.0 & 12 & 0.7 & 10.5 \\
\hline
\label{tab:properties}
\end{tabular}
\end{table}

It is interesting to note that two relatively close pairs of galaxies are
observed in the AXAF field.  EISU28/EISU12 and EISB07/EISB02 are
separated of only 8.7 and 3.1 proper Mpc, respectively.
Given the small number of objects in the present sample any
statistical conclusion is obviously impossible, but it appears natural
to link  the occurrence of the two pairs to the redshift
``spikes'' observed by Steidel et al. (1998) at $z\sim3$.
Future observations of the remaining high-z galaxy candidates 
and the extension of the surveyed area (see {\tt
http://www.eso.org/science/eis/}) will make it possible
 to address also this issue on a more
quantitative basis.
 
\begin{acknowledgements}
We warmly thank F. Comeron, 
R.Gilmozzi, P.Rosati and all the FORS-1/ISAAC SV Team and the EIS Team
for making possible the present observations.  AAS acknowledges
generous financial support form the Royal Society.  SA has been
supported during this work by a Marie Curie Grant Fellowship.  This
work has been conducted with partial support by the TMR programme
Formation and Evolution of Galaxies set up by the European Community
under the contract FMRX-CT96-0086.
\end{acknowledgements}

\end{document}